\documentclass[12pt]{article}
\begin{document}
\begin{center}
{\Large \bf Final state interaction and $\Delta I=1/2 $ rule  }\\
\vspace{5mm}
{{\bf E.P. Shabalin}\footnote{{\it E-mail address:} shabalin@itep.ru
(E.Shabalin)}} \\{\it Institute for Theoretical and Experimental Physics,
B.Cheremushkinskaya, 25, 117218, Moscow, Russia }\\ \end{center}
\vspace{1cm}
\noindent{\bf Abstract}\\
{\small Contrary to wide-spread opinion that the
final state interaction (FSI) enhances the amplitude $<2\pi;I=0|K^0>$, we
argue that FSI does not increase the absolute value of this
amplitude.}\\

\vspace{3mm}

\vspace{1cm}

The essential progress in understanding the nature of the $\Delta I=1/2$
rule in $K\to 2\pi$ decays was achieved in the paper \cite{1}, where
the authors had found a considerable increase of contribution of the
operators containing a product of the left-handed and right-handed quark
currents generated by the diagrams called later the penguin ones.  But for
a quantitative agreement with the experimental data, a search for some
additional enhancement of the $<2\pi;I=0|K^0>$ amplitude produced by
long-distance effects was utterly desirable. A necessity of additional
enhancement of this amplitude due to long-distance strong interactions was
also noted later in \cite{2}.

The attempts to take into account the long-distance effects were undertaken
in \cite{3} - \cite{14}.

In \cite{3}, the necessary increase of the amplitude $<2\pi;I=0|K^0>$ was
associated with 1/N corrections calculated within the large-N approach
(N being the number of colours).

In \cite{4}, \cite{5}, the strengthening of the $<2\pi;I=0|K>$ amplitude
arised due to a small mass of the intermediate scalar $\sigma$ meson.

One more mechanism of enhancement of the $<2\pi;I=0|K^0>$ amplitude was
ascribed to the final state interaction of the pions \cite{6} - \cite{14}.
But as it will be shown in present paper,
unitarization of the $ K\to
2\pi$ amplitude in presence of FSI leads to the opposite effect: a
decrease of the $<2\pi;I=0|K^0>$ amplitude.

We exploit the technique based on the effective $\Delta S =1$ non-leptonic
Lagrangian \cite{1} \begin{equation} L^{\rm{weak}} =\sqrt{2}G_F \sin
\theta_C \cos \theta_C \sum_i c_i O_i.
\end{equation}
Here $O_i$ are the four-quark operators and $c_i$ are the Wilson
coefficients calculated taking into account renormalization effect
produced by strong quark-gluon interaction at short distances.
Using also the recipe for bosonization of the diquark compositions
proposed in \cite{2}, one obtains the following result:
\begin{equation}
<\pi^+(p_+),\pi^-(p_-);I=0|K^0(q)>=\kappa(q^2-p^2_-),
\end{equation}
where
$\kappa$ is a function of $G_F, F_{\pi}, \theta_C$ and some combination
of $c_i$. The numerical values of $\kappa$ obtained in \cite{1} and
\cite{2} turned out
to be insufficient for a reproduction of the observed magnitude of the
$<2\pi;I=0|K^0>$ amplitude.

Could a rescattering the final pions occuring at long distances change
the situation? To answer this question, we consider at first the elastic
$\pi\pi$ scatterig itself.\\
\noindent{\bf The elastic $\pi\pi$ scattering.} \\
The general form of the amplitude of elastic $\pi\pi $ scattering is
\begin{equation}
T=<\pi_k(p'_1) \pi_l(p'_2)|\pi_i(p_1)\pi_j(p_2)> =A\delta_{ij} \delta_{kl}
+B\delta_{ik}\delta_{jl} + C\delta_{il}\delta_{jk},
\end{equation}
where $k,l,i,j$ are the isotopical indices and $A,B,C$ are the functions
of $s=(p_1+p_2)^2$, $t=(p_1-p'_1)^2$ ,$u=(p_1-p'_2)^2$.

The amplitudes with the fixed isospin $I=0,1,2$ are
\begin{equation}
T^{(0)}=3A+B+C, \qquad T^{(1)}=B-C, \qquad T^{(2)}=B+C .
\end{equation}
To understand the problems arising in description of $\pi\pi$ scattering in
the framework of field theory, let's consider the simplest chiral $\sigma$
model, where
\begin{equation}
A^{\rm{tree}}=
\frac{g^2_{\sigma\pi\pi}}{m^2_{\sigma}-s}-\frac{g^2_{\sigma\pi\pi}}
{m^2_{\sigma}-m^2_{\pi}}
=\frac{g^2_{\sigma\pi\pi}}{m^2_{\sigma}-m^2{\pi}}\cdot
\frac{s-m^2_{\pi}}{m^2_{\sigma}-s}
\end{equation}
and $B$ and $C$ are obtained from $A$ by replacement $s\to t$ and $s\to
u$, respectively.

It follows from Eqs.(4) and (5), that the isosinglet amplitude
$T^{(0)}_{\rm{tree}}$ is a sum of the resonance part \begin{equation}
A_{\rm{Res}}^{\rm{tree}}=3A^{\rm{tree}} \end{equation} and the
potential part \begin{equation} A^{\rm{tree}}_{\rm{Pot}}=
B^{\rm{tree}}+C^{\rm{tree}}. \end{equation} The resonance part must be
unitarized summing up the chains of pion loops, that is , taking into account the repeated
rescattering of the final pions.

At the one loop oder
\begin{equation}
A^{\rm{one-loop}}_{\rm{Res}}=A^{\rm{tree}}_{\rm{Res}}(1+\Re\Pi_R+i
\Im\Pi)=A^{\rm{tree}}_{\rm{Res}}(1+ \Re\Pi_R+
i\frac{A^{\rm{tree}}_{\rm{Res}}\sqrt{1-4m^2_{\pi}/s}}{16\pi}),
\end{equation}
where $\Re \Pi_R$ is the renormalized real part of the closed pion loop
\cite{15}
\begin{equation}
\Re\Pi_R(s)=\Re\Pi(s)-\Re\Pi(m^2_{\sigma}) -\frac{\partial
\Re\Pi(s)}{\partial s}|_{s=m^2_{\sigma}}(s-m^2_{\sigma}).
\end{equation}
The last two terms in r.h.s. of this equation are absorbed in
renormalization of the resonance mass and coupling constant
$g_{\sigma\pi\pi}$.
Though $\Re\Pi_R(s)$ can be calculated to leading order in
$g_{\sigma\pi\pi}$ \cite{16}, in view of very big value of this constant
such a calculation does not give a proper estimate of $\Re \Pi _R(s)$. It
will be explained below, how to get a reliable magnitude of $\Re\Pi_R(s)$.

The unitarized expression for $A_{\rm{Res}}$ is \footnote{Strictly
speaking, the $4\pi$ intermediate state brings a correction in Eq.(10).
But its contribution to $\Im \Pi(s)$ is equal to zero because
$4m_{\pi}>m_K$. As for $\Re \Pi_R(s)$, in our approach, all separate
contributions to it will be taken into account phenomenologically
introducing a form factor, see below Eq.(18).}

\begin{equation}
A^{\rm{unitar}}_{\rm{Res}}=\frac{A^{\rm{tree}}_{\rm{Res}}(s)}{1-\Re\Pi_R(s)-i\Im\Pi_{\rm{Res}}}=
\frac{A^{\rm{tree}}_{\rm{Res}}(s)}{1-\Re\Pi_R(s)}\cdot \frac{1}{1-i\tan
\delta_{\rm{Res}}},
\end{equation}
where
\begin{equation}
\tan \delta_{\rm{Res}}=\frac{A^{\rm{tree}}_{\rm{Res}}(s)
\sqrt{1-4m^2_{\pi}/s}}{16\pi(1-\Re\Pi_R(s))}.
\end{equation}
The Eq.(10) may be rewritten in the form
\begin{equation}
A^{\rm{unitar}}_{\rm{Res}}=\frac{16\pi \sin \delta_{\rm{Res}}
e^{i\delta_{\rm{Res}}}}{\sqrt{1- 4m^2_{\pi}/s}}, \end{equation} leading to
the cross section \begin{equation} \sigma_{\rm{Res}}=\frac{4\pi
\sin^2\delta_{\rm{Res}}}{k^2}, \qquad k=\frac{\sqrt{s}}{2} \cdot
\sqrt{1-4m^2_{\pi}/s}.  \end{equation}

Of course, the amplitude $T^{(0)}$ must be unitarized including the
potential part $B+C$ too. But if this potential part is considerably
smaller than the resonance one, the effect of FSI can be estimated
roughly from $A^{\rm{unitar}}_{\rm{Res}}$. To
understand what gives the unitarization of $A^{\rm{tree}}_{\rm{Pot}}$
, we use the form of the $S$ matrix of elastic
scattering with the total phase shift as a
sum of the phase shifts produced by separate mechanisms of scattering
\cite{17}. In other words, if there is a number of resonances and if, in
addition, there is potential scattering, the matrix $S$ looks as
\begin{equation} S=e^{2i\delta_{\rm{Res1}}}
e^{2i\delta_{\rm{Res2}}}...e^{2i\delta_{\rm{Pot}}}. \end{equation} Then,
in terms of \begin{equation} \delta_{\rm{Res}}=\sum_j \delta_{\rm{Resj}}
\quad\mbox{and} \quad
\delta_{\rm{tot}}=\delta_{\rm{Res}}+\delta_{\rm{Pot}}, \end{equation}
\begin{equation} A^{\rm{unitar}}=\frac{16\pi}{\sqrt{1-4m^2_{\pi}/s}}\sin
\delta_{\rm{tot}}e^{i\delta_{\rm{tot}}}
\end{equation}
or
\begin{equation}
A^{\rm{unitar}}=\frac{16\pi}{\sqrt{1-4m^2_{\pi}/s}}(\sin \delta_{\rm{Res}}
\cos \delta_{\rm{Pot}}+\sin \delta_{\rm{Pot}} \cos
\delta_{\rm{Res}})e^{i\delta_{\rm{tot}}}. \end{equation}

The phase shifts $\delta_{\rm{Res}}$ and $\delta_{\rm{Pot}}$ can be taken
from \cite{18}, where the Resonance Chiral Theory of $\pi \pi$ Scattering
was elaborated. This model incorporates two $\sigma$ mesons, $f_0(980)$,
$\rho(750)$ and $f_2(1270)$. In addition, some phenomenological form
factors were introduced in the vertices $\sigma\pi\pi, \rho\pi\pi,
f_2\pi\pi$.  Their appearence follows in the field theory from the result
(10), according to which the effect of $\Re \Pi_R(s)$ may be incorporated
in $g^2_{\sigma \pi\pi}(s)$, where \begin{equation} g^2_{\sigma
\pi\pi}(s)=\frac{g^2_{\sigma \pi\pi}}{1-\Re \Pi_R(s)}= g^2_{\sigma
\pi\pi}F(s).  \end{equation} The model gives a quite satisfactory
description of the observed behavior of the phase shifts $\delta^0_0(s)$,
$\delta^2_0(s)$, $\delta^1_1(s)$ in the range $4m^2_{\pi}\le s \le 1
\mbox{GeV}^2$. The phase shifts $\delta^0_2(s)$ and $\delta^2_2$ turn out
to be consistent with the results obtained using the Roy's dispersion
relations.

Using the parameters found in \cite{18}, one obtains
$F(\sqrt{s}=m_K)=0.894$ and
\begin{equation}
\Re \Pi_R(s=m^2_K)=-0.12.
\end{equation}
For $\sqrt{s}=m_K$, the phase shifts obtained in \cite{18} are
\begin{equation}
\delta_{\rm{Res}}=46.71^{\circ}, \qquad \delta_{\rm{Pot}}=-9.40^{\circ}.
\end{equation}
Then
\begin{equation}
\frac{A^{\rm{unitar}}_{\rm{Pot}}}{A^{\rm{unitar}}_{\rm{Res}}}=\frac{\sin
\delta_{\rm{Pot}} \cos \delta_{\rm{Res}}}{\sin \delta_{\rm{Res}} \cos
\delta_{\rm{Pot}}}=-0.156.  \end{equation} Therefore, the amplitude
$A^{\rm{unitar}}_{\rm{Pot}}$ is small and may be neglected in a rough
estimate of FSI effect.

A value of the total tree amplitude produced by $\sigma$ exchange,
calculated using the parameters found in \cite{18} is
\begin{equation}
A^{\rm{tree}}(s=m^2_K)=55.22.
\end{equation}
The unitarization of this amplitude gives according to Eqs.(17) and (20)
\begin{equation}
|A^{\rm{unitar}}(s=m^2_K)|=36.95.
\end{equation}
Therefore, the unitarization decreases the tree amplitude by 1.49 times!
The analogous effect of FSI takes place in the $K \to 2\pi$ decay.\\
\noindent{\bf FSI in $K^0 \to 2\pi$ decay.}\\
Basing on the result (21), we shall estimate effects of FSI
in the $K \to 2\pi$ amplitude, taking into account only
the resonance rescattering effect.  Then, in one loop approximation,
the amplitude (2) is \begin{equation}
\begin{array}{lll} <\pi^+ (p_+)
\pi^-(p_-);I=0|K^0(q)>^{\rm{one-loop}}_{\rm{Res}}=    \\ \kappa
\left[(q^2-p^2_-)+ \frac{A^{\rm{tree}}_{\rm{Res}}(q^2)}{(2\pi)^4
i}\int\frac{(q^2-p^2)d^n
p}{[(p-q)^2-m^2_{\pi}][p^2-m^2_{\pi}]}+i\frac{A^{\rm{tree}}_{\rm{Res}}(q^2)}{16\pi}
(q^2-p^2_-)\sqrt{1-
4m^2_{\pi}/q^2}\right].
\end{array}
\end{equation}

In the t'Hooft-Veltman scheme of dimensional regularization \cite{19}
\begin{equation}
\begin{array}{lll}
\frac{1}{(2\pi)^4i}\int\frac{d^n p}{[(p-q)^2-m^2][p^2-m^2]}=
\frac{1}{16\pi^2}\left(\ln\frac{M^2_0}{m^2}+2+  \sqrt{1-4m^2/q^2} \ln
\frac{1- \sqrt{1-4m^2/q^2}}{1+\sqrt{1-4m^2/q^2}} \right);\\
\frac{1}{(2\pi)^4}\int \frac{p^2 d^np}{[(p-q)^2-m^2][p^2-m^2]}=
\frac{m^2}{16\pi^2}\left(2\ln\frac{M^2_0}{m^2}+3+\sqrt{1-4m^2/q^2}\ln
\frac{1-\sqrt{1-4m^2/q^2}}{1+\sqrt{1-4m^2/q^2}} \right)\\
M_0\to \infty .
\end{array}
\end{equation}
After renormalization excluding the parts of these integrals independent
of the external momentum, we come to
\begin{equation}
\begin{array}{lll}
<\pi^+ \pi^-|K^0(q)>^{\rm{one-loop}}_{\rm{on-mass-shell}}= \\
=\kappa
(m^2_K-m^2_{\pi})\left[1+\frac{A^{\rm{tree}}_{\rm{Res}}(s)}{16\pi^2}\sqrt{1-
4m^2_{\pi}/s} \ln \frac{1-\sqrt{1-4m^2_{\pi}/s}}{1+\sqrt{1-4m^2_{\pi}/s}} +
i\frac{A^{\rm{tree}}_{\rm{Res}}(s)}{16\pi}\sqrt{1-4m^2_{\pi}/s} \right].
\end{array}
\end{equation}
This result agrees with the Cabibbo-Gell-Mann theorem \cite{20}, according
to which the $K\to 2\pi$ amplitude vanishes in the limit of exact
$SU(3)$ symmetry.
From Eq.(26) in the leading order of perturbation theory one has
\begin{equation} \Re\Pi_R(s)
=\frac{A^{\rm{tree}}_{\rm{Res}}(s)}{16\pi^2}
\sqrt{1-4m^2_{\pi}/s}\ln\frac{1-\sqrt{1-4m^2_{\pi}/s}}{1+\sqrt{1-
4m^2_{\pi}/s}}.
\end{equation}
But, as it was noted above, the perturbation theory does not give a
reliable value of $\mbox{Re} \Pi_R(s)$ and for its estimate some more
complicated procedure (described above) must be applied.

The unitarization of the amplitude (26) done
in accordance with the prescription (10) leads to the result
\begin{equation}
|<\pi\pi;I=0|K^0(q^2=m^2_K)>|_{\rm{Res}}=\kappa
(m^2_K-m^2_{\pi})\frac{\cos\delta_{\rm{Res}}}{1-\Re
\Pi_R(m^2_K) }. \end{equation}
This part yields 0.61 of a value of the initial amplitude (2) and the part
connected with the potential rescattering, being negative, can not change
the conclusion that FSI diminishes the tree amplitude.

The influence of FSI on the $K^0\to 2\pi$ decay was studied in the
framework of $\sigma$ model in the papers \cite{21}. In these papers, the
authors, however, put $\Re \Pi=0$. Then
$$
A^{\rm{unitar}}=A^{\rm{tree}}/(1-i\Im \Pi)
$$
and this formula was used by them to estimate the FSI effects in the $K^0
\to 2\pi$ decay. But earlier the same authors had found that $\Re \Pi \ne
0$ \cite{16}. In this case, the unitarization leads to
$A^{\rm{unitar}}$ for the elastic $\pi\pi$ scattering given in
Eq.(10)  and to $A^{\rm{unitar}}_{\rm{Res}}(K\to 2\pi; I=0)$
in Eq.(28). As it is seen from Eq.(28), FSI could increase
or diminish the $K\to 2\pi$ amplitude depending on relative magnitudes of
$\cos \delta$ and $(1-\Re \Pi_R)$ . We have shown that
$\cos\delta/(1-\Re\Pi_R)<1$, that allows us to affirm that FSI
diminishes the isosinglet part of the $K \to 2\pi$ amplitude.

\noindent{\bf Conclusion.}\\
We have not found an enhancement of the amplitude $<\pi\pi;I=0|K^0>$ due to
final state interaction of pions. On the contrary, our analysis has shown
that FSI diminishes this amplitude. Hence, FSI is not at all the mechanism
bringing us nearer to explanation of the $\Delta I=1/2$ rule in the $K\to
2\pi$ decay. As for the results \cite{3} - \cite{5}, obtained without
unitarization of the $K\to 2\pi$ amplitude, they ought to be
reconsidered.\\

\vspace{1cm}\noindent{\bf Acknowledgements}\\

\vspace{5mm}

I am very grateful to Yu.A.Simonov for discussions and comments concerning
this letter.

\end{document}